\documentclass{ws-procs975x65}

\begin{document}

\title{Detecting LISA sources using time-frequency techniques}

\author{JONATHAN R GAIR}

\address{Institute of Astronomy, University of Cambridge, Cambridge, CB3 0HA, UK\\
\email{jgair@ast.cam.ac.uk}}

\author{GARETH JONES}

\address{Cardiff School of Physics and Astronomy, Cardiff University, Queens Buildings,
The Parade, Cardiff, CF24 3AA, UK\\
\email{Gareth.Jones@astro.cf.ac.uk}}


\bodymatter

\section{Introduction}\label{intro}
The LISA data stream will contain many gravitational wave (GW) signals from different types of source, overlapping in time and frequency. We expect to detect signals from compact binaries (composed of white dwarfs (WDs) or neutron stars (NSs)), in the nearby Universe. At low frequencies these will form a confusion foreground, but we also hope to individually resolve $\sim10,000$ of these sources\cite{farmer} at high frequencies. LISA will also detect 1-10 signals per year\cite{sesana05} from the merger of supermassive black holes (SMBHs) of appropriate mass ($\sim10^5 M_{\odot}$ -- $10^7 M_{\odot}$). Thirdly, LISA should detect GWs from extreme mass ratio inspirals (EMRIs) --- the inspiral of a compact object (a WD, NS or BH) into a SMBH in the centre of a galaxy. The astrophysical rate is very uncertain, but LISA could resolve as many as several hundred EMRIs \cite{gair04} and may also see a confusion background from distant events\cite{bc04}.

The development of techniques to analyze LISA data is the subject of much current research. One promising approach is to use Markov Chain Monte Carlo (MCMC) methods. These have proven effective for detecting compact binaries\cite{cornish05}, SMBH mergers\cite{cornishporter06} and for the detection of a single simplified EMRI signal\cite{stroeer06emri}. Although MCMC techniques can be used to fit simultaneously for many signals of several types, it is not yet clear whether this will be practical for the EMRI search. This is because of the high computational cost associated with constructing sufficiently accurate EMRI waveform templates, even when using kludge models\cite{gair04}. It may therefore by impractical to use MCMC for the EMRI search unless some advance estimate has been made of the source parameters. One alternative approach to LISA data analysis is to use time-frequency (t-f) techniques. These could be used to estimate the parameters of the loudest EMRIs in the LISA data stream and for the detection of unexpected GW events. A t-f analysis will consist of two stages --- detection of a source in the data and parameter estimation for that source. 

\section{Source Detection}
We consider a simplified model of the LISA data stream in which there is a single source embedded in instrumental noise. We divide the data stream into $M$ segments of length $T$, carry out a Fourier transform on each segment and hence construct a spectrogram of the data, $S^0$, with power $P^0_{i,j}$ in pixel $(i, j)$. We then search this spectrogram for features. The simplest technique is to look for individual pixels that are unusually bright, i.e., with $P^0_{i,j} > \eta$, for some suitably chosen threshold $\eta$. To improve the performance, we generate and search a sequence of binned spectrograms, $S^k$, in which the power in pixel $(i,j)$ is defined to be
\begin{equation}
P^k_{i,j} = \frac{1}{n_k \times l_k} \sum_{a=0}^{n_k-1} \sum_{b=0}^{l_k-1} P^0_{i+a,j+b}.
\end{equation}
Using bins of the form $n_k=2^p$, $l_k=2^q$, for all possible $p$ and $q$, a segment length $T=2^{20}$s, and assuming a $3$ year LISA mission, this simple excess power search has a reach of $\sim2.5$Gpc for a typical EMRI event (we take the reach to be the distance at which the detection rate is only $20\%$ for a search false alarm probability of $10\%$).  The range is somewhat higher for EMRIs on nearly circular orbits. This method and these results are described in Wen \& Gair 2005\cite{wengair05} and Gair \& Wen 2005\cite{gairwen05}.

A more sophisticated technique is to look for clusters of bright pixels. One algorithm is the Hierarchical Algorithm for Clusters and Ridges (HACR). This involves identifying {\it black pixels} with $P_{i,j} > \eta_{up}$, and then counting the number of {\it grey pixels} with $P_{i,j} > \eta_{low}$ ($< \eta_{up}$) that are connected to the black pixel. If the number of pixels in the cluster, $N_p$, exceeds a threshold, $N_c$, then the cluster constitutes a detection. The three thresholds can be tuned to make the search sensitive to a particular source or chosen to make the search generally sensitive to a variety of source types. After tuning, HACR has a detection rate $10-15\%$ higher than the simple excess power search at fixed overall false alarm probability for a typical EMRI. This represents a significant improvement in LISA event rate. The HACR search is described in more detail in Gair \& Jones 2006\cite{gairjones06}. HACR can also detect SMBH mergers at redshift up to $\sim 3.5$ and compact binaries at up to $\sim 12$kpc\cite{gairjones06}.

\section{Parameter Extraction}
Once a source has been identified in the data, we would like to estimate its parameters to 
allow a targeted follow up with matched filtering. The time-frequency structure of an event 
tells us about the type of signal --- a WD-WD binary is almost monochromatic (the track is 
therefore long in time but narrow in frequency), while EMRI and SMBH merger signals ``chirp'' 
over time. EMRIs chirp slowly and are likely to be on eccentric orbits, indicated by the 
presence of several tracks at different frequencies that evolve in a similar fashion. By 
contrast, SMBH mergers are likely to be circular and evolve much more rapidly. The time, 
central frequency, frequency derivatives and power profile of an event can all be extracted 
from a t-f map and provide information on the system, as does the bin size used to generate 
the spectrogram in which the detection is made. If multiple tracks can be associated with the 
same event 
we get this information for each track. 
The shape of the boundary of a track provides a way to distinguish a single event from two crossing tracks or a noise burst. The shape parameters (curvature, area, perimeter), skeleton and convex hull of a cluster provide further information\cite{russ02}. This information can be extracted directly from clusters identified by HACR (see discussion in Gair \& Jones 2006\cite{gairjones06} and Gair \& Jones 2007 in prep.). The excess power search identifies individual pixels only, so this search must be followed by a second track identification search before information can be extracted\cite{lisa6proc}.

\section{Application to LISA Data Analysis}
Time-frequency searches of the nature described here could play a useful role in the LISA data analysis pipeline. These methods should be able to detect the loudest events in the LISA data stream at much lower computational cost than matched filtering searches. They also provide a method to find unexpected sources in the LISA data, since they do not rely on the observer having a model of the source. The main issue that will limit the sensitivity of time-frequency techniques is source confusion. The analyses described here have considered the detection of single isolated events, which is not the situation we expect for LISA. To deal with confusion, we could apply t-f techniques only to analyze a ``cleaned'' spectrogram, i.e., with the loudest recognizable events extracted as well as possible by other techniques. This could find events missed at the first stage of the analysis, but the effect of cleaning must be carefully explored. Alternatively, we can use percolation techniques --- set a high threshold and gradually reduce it until a track appears. We can then extract this loudest event before lowering the threshold further to find the next event etc. This approach will be examined further in the future. Although our focus has been on LISA, the methods discussed here could also be applied to searches of Advanced LIGO data, e.g., for detection of intermediate mass ratio inspiral sources.

\section*{Acknowledgments}
This work was supported by St. Catharine's College, Cambridge (JG) and by the School of Physics and Astronomy, Cardiff University (GJ).


\end{document}